# Pattern formation during nonequilibrium crystallization by classical-density-functional-based approach

Kun Wang[1*], Shifang Xiao[2], Jun Chen[3], Wangyu Hu[1 †]

[1] College of Materials Science and Engineering, Hunan University, Changsha 410082, China
[2] Department of Applied Physics, Hunan University, Changsha 410082, China
[3] Institute of Applied Physics and Computational Mathematics, Beijing 100088, China

Solidification pattern during nonequilibrium crystallization is among the most important microstructures in the nature and technical realms. Phase field crystal (PFC) model could simulate the pattern formation during equilibrium crystallization at atom scale, but cannot grasp the nonequilibrium ones due to the absence of proper elastic-relaxation time scale. In this work, we propose a minimal classical-density-functional-theory-based model for crystal growth in supercooled liquid. Growth front nucleation (GFN) and various nonequilibrium patterns, including the faceting growth, spherulite, dendrite and the columnar-to-equiaxed transition (CET) among others, are grasped at atom scale. It is amazing that, except for undercooling and seed spacing, seed distribution is key factor that determines the CET. Overall, two-stage growth process, i.e., the diffusion-controlled growth and the GFN-dominated growth, are identified. But, compared with the second stage, the first stage becomes too short to be noticed under the high undercooling. The distinct feature for the second stage is the dramatic increments of dislocations, which explains the amorphous nucleation precursor in the supercooled liquid. Transition time between the two stages at different undercooling are investigated. Crystal growth of BCC structure further confirms our conclusions.

## I. INTRODUCTION

Nonequilibrium crystallization in undercooled liquid is frequently encountered in the nature (e.g., snowflakes and minerals) and technical realms, for example, the dendrites in traditional as-cast materials or additively manufactured parts[1], and spherulites in Se[2], polymers[3] and so on. Spatially heterogenous dynamics in the supercooled liquid have been well established experimentally [4, 5], which have numerous consequence on the transport properties. The most important transport properties relevant to the crystallization are the shear viscosity and the molecular mobilities determined by the translational and rotational diffusion coefficients[6]. The latter ones directly control detailed manners of how molecules attach and align with the growing crystal. Particularly, the spherulitic growth patterns are results of competitions between the translational and rotational motions, which have ever puzzled researchers in the related field for a long time [3, 7] and later been clarified by Gránásy et. al.[8, 9] using phase field model extended via allowing for evolutions of local crystal orientations, termed orientation-field-based phase field (OFPF) model. A key result of the competitions is emergence of *growth front nucleation* (GFN) which is a new formation mechanism for polycrystalline in contrast to the traditional

---

one through impinging among growing single crystals [10]. Another important phenomenon relevant to the nonequilibrium crystallization is the columnar-to-equiaxed transition (CET) which has many technical applications. The CET is captured fairly well by a phenomenological model of Hunt [11], however, in terms of parameters (such as temperature gradient and velocity of the growth front) difficult to be quantified. The OFPF simulations indicates that foreign-particle induced nucleation under the high undercooling plays the key role for the CET[10]. Despite of the great achievements in understanding of the nonequilibrium crystallizations, present knowledge based on the OFPF model rely on the assumption that crystals free to change its local orientation to lower the free energy, which is not entirely true for crystals and sometimes result in predictions that are qualitatively wrong. For example, rotation and shrinking of circular grain embedded into an infinite crystal cannot be correctly predicted by the OFPF model due to geometrical constraints on the dislocations (See the review paper [10]). This indicates that microstructures at atom scale is important. Indeed, Tegze et al [12] found that the diffusion-controlled (or slow) growth mode and the diffusionless (or fast) steady growth mode during equilibrium crystallizations have distinctly different interface structures. The interface for the former is fairly thin and faceted, while the latter one extends to several atomic layers and shows rounded corners. As results, different growth morphonology emerge. Actually, the evolution of the local crystal orientation, the interface energy and its anisotropy are governed by the microstructure and the latter is determined by the undercooling of solidification system. Thereby, complete understanding of the nonequilibrium crystallizations requires us to treat the crystal growth at a more fundamental level, i.e., in terms of the microstructures at atom level, as well as its kinetics. However, simulation of the crystallization process by present atomic simulation methods, such as classical molecular dynamics, remains a great challenge at present due to the limitation of spatial and temporal scale. The classical-density-functional-theory-based phase field crystal (PFC) model, firstly proposed by Elder et. al. [13], is a promising candidate and has been used to predict the symbiosis of faceting with growth mode selection via appending a colored Gaussian noise term in the motion equation of the reduced atom density [12]. Amorphous nucleation precursor in the supercooled liquid is successfully caught by the PFC model [14], but the dendrite growth for pure elements only exists in the solid-liquid coexistence region of the PFC phase diagram [12, 15].

Recently, Podmaniczky et al[16] successfully grasped the GFN as well as the spherulite in the undercooled system using a hydrodynamics-coupled phase field crystal (HPFC) model. Although other nonequilibrium phenomenon during the crystallizations, such as the dendritic growth as and CET, are not shown in their results (perhaps due to the low computational efficiency of the model), it implies that certain "dynamics" contained in the hydrodynamics, but absent in the PFC model, is crucial for describing the nonequilibrium phenomenon during crystallizations. Interestingly, other kind of hydrodynamics-coupled PFC model could solve the problem of elastic phonon relaxations absent in original PFC dynamics [17, 18]. Similar dynamics is also observed in a PFC model extended through coupling with dislocation stress field at elastic equilibrium states[19, 20]. These breakouts have greatly promoted our understandings on the nonequilibrium phenomenon from macro/mesoscopic levels, but the key ingredient missing in the original PFC model at atom or lattice level is still unknown.

In this work, we make an attempt to understanding the nonequilibrium crystallizations using a minimal extended PFC model. In contrast to the existing models extended through incorporating other known theories that usually contains concepts not existing in the original model and thus requires establishing complex coupling relationship between them, our model is completely based on the classical functional theory. The key ingredient of our model is considering the elastic relaxations through allowing for additional lattice rotations. Numerical results show the ability to capture the GFN and various nonequilibrium patterns, including the faceting growth, spherulite, dendrite and the CET among others at atom scale, unprecedently. Specially, the role of crystallite interaction as well as the undercooling on the crystal growth is investigated. It is found that the nonequilibrium pattern does not only depend on the undercooling, lattice symmetry and seed spacing, but also on the distribution and type of the crystallites where multi-crystallite interaction plays a key role. Our results reveal a two-stage growth process. The first stage is the diffusion-controlled anisotropic growth and the later stage is the GFN-dominated growth. The transition time between them relies on the undercooling by a power law. Although we mainly discuss the crystal growth of a 2D model crystal (directly relevant to 2D colloids, surfactant monolayers, island formation on substrates, polymer films and so on), the conclusions to be addressed is rather general. Besides, crystal growth of BCC structure is investigated, which further confirms our conclusions.

## II. CLASSICAL-DENSITY-FUNCTIONAL-THEORY EXTENDED PFC MODEL

In classical density functional theory[21, 22], Helmholtz free energy ($\mathcal{F}$) of solid-liquid coexistent system could be expressed as a summation of noninteractive part ($\mathcal{F}_{id}$), excess free energy ($\mathcal{F}_{ex}$) and the contribution from external fields, which is a functional of one-particle density ($\rho(\mathbf{r})$). Particularly, $\mathcal{F}_{id}$ is in essence the free energy of ideal gas, i.e.,

$$\mathcal{F}_{id} = k_B T \int \rho[\ln(\Lambda^d \rho) - 1] d\mathbf{r}, \tag{1}$$

where $k_B$ is Boltzmann constant, $T$ is temperature, $d$ is dimension of the system and $\Lambda$ is the thermal de Broglie wavelength. The exact expression for $\mathcal{F}_{ex}$ depends on the detailed interactions of the system, which is usually unknown a prior. Under the Ramakrishnan-Yussouff approximation[22], the excess free energy for pure element could be expressed by

$$\mathcal{F}_{ex}(T, [\rho(\mathbf{r})]) = -\frac{1}{2} k_B T \int \int d\mathbf{r}_1 d\mathbf{r}_2 \Delta\rho(\mathbf{r}_1) C^{(2)}(\mathbf{r}_1, \mathbf{r}_2) \Delta\rho(\mathbf{r}_2), \tag{2}$$

where $C^{(2)}$ is direct pair-correlation function and $\Delta\rho(\mathbf{r}_j) = \rho(\mathbf{r}) - \rho_l$, $\rho_l$ is the number density of the liquid phase. The PFC model further estimates $C^{(2)}$ by gradient expansions to the fourth order[21], i.e.,

$$C^{(2)} \approx C_0^{(2)} - C_2^{(2)} \nabla_\mathbf{r}^2 + C_4^{(2)} \nabla_\mathbf{r}^4, \tag{3}$$

where $C_k^{(2)}$ ($k = 0, 2, 4$) are expansion coefficients and the gradients of odd order vanish due to $C^{(2)}(\mathbf{r}) = C^{(2)}(-\mathbf{r})$. Define the dimensionless reduced density

$$\psi(\mathbf{r}) = (\rho(\mathbf{r}) - \bar{\rho})/\bar{\rho}, \tag{4}$$

where $\bar{\rho}$ is the average number density. Through expanding the integrand of $\mathcal{F}_{id}$ around $\psi = 0$ and combining with Eqs. (2,3), the free energy as a functional of $\psi(\mathbf{r})$ is obtained. Various variants could be formulated, but they are only differed by the coefficients ahead of each term. Swift-Hohenberg model[13, 23] is one of the variants, whose dimensionless free energy functional reads

$$\mathcal{F} = \int d\mathbf{x} \left\{ \frac{\psi}{2} [-\epsilon + (1 + \nabla^2)^2] \psi + \frac{\psi^4}{4} \right\}, \tag{5}$$

where $\psi$ is dimensionless local atomic density and $\epsilon$ is a small quantity proportional to the undercooling extent of the system. Fourier expansion of $\psi$, to the principal reciprocal lattice vectors (RLVs) of considered structure, gives

$$\psi(\mathbf{x}) = \bar{\psi} + \sum_{j=1}^{N} \eta_j \exp(i\mathbf{K}_j \cdot \mathbf{x}) + c.c., \tag{6}$$

where $c.c.$ denotes complex conjugate of the summation term, $\bar{\psi}$ is average reduced density and $\eta_j$ is a complex amplitude of the density wave with a RLV of $\mathbf{K}_j$. The summation runs over all (totally $N$) principal RLVs. Hereafter, we will take 2D hexagonal lattice for example and the method can be also applied for other lattices (Specially, the result for BCC lattice is given in the Appendix). Initial RLVs of the hexagonal lattice are $\mathbf{K}_1 = k_0(-\sqrt{3}/2, -1/2)$, $\mathbf{K}_2 = k_0(0,1)$, $\mathbf{K}_3 = k_0(\sqrt{3}/2, -1/2)$, where $k_0 = 1$ at equilibrium (reference) states.

We firstly follow the variational approach[24, 25] to derive the coarse-grained model and then extend the resulting model through treating the elastic relaxations within the framework of classical density functional theory. Through substituting Eq. (6) into the integrand of Eq. (5) and integrating the result over a unit cell, the coarse-grained free energy density can be obtained by dividing the result with the cell volume. Because the average density, amplitudes and wave vectors (or RLVs) vary slowly in space in comparison with the density wave itself, $\bar{\psi}$, $\eta_j$ and $\mathbf{K}_j$ could be viewed as functions of a slowly varying spatial variable and thus remain constant during the cell integration. The resulting coarse-grained free energy functional is

$$\mathcal{F}^{cg} = \int d\mathbf{x} \left\{ \frac{1}{2}[-\epsilon + 1]\bar{\psi}^2 + \frac{\bar{\psi}^4}{4} - (\epsilon - 3\bar{\psi}^2)A^2 + 3A^4 + \sum_{j=1}^{3}|\mathcal{G}_j \eta_j|^2 - \frac{3}{2}\sum_{j=1}^{3}|\eta_j|^4 + 6\bar{\psi}\left(\prod_{j=1}^{3}\eta_j + \prod_{j=1}^{3}\eta_j^*\right)\right\}$$

(7)

where $\mathcal{G}_j = \nabla^2 + 2i\mathbf{K}_j \cdot \nabla$, $A^2 = \sum_{j=1}^{N}|\eta_j|^2$. Besides, the higher order derivative terms of $\bar{\psi}$ have been dropped since these terms would lead to numerical instabilities[24] and can be self-consistently eliminated through a convolution operator[26].

Simulating dendrite growth requires both "slow" diffusion dynamics and "fast" diffusionless dynamics, i.e., elastic relaxations. The "slow" diffusion dynamics has been well developed in terms of evolutions of complex amplitudes and average density in literatures [24, 25, 27]. Namely, the complex amplitudes obey nonconserved dissipative dynamics, while the average density follows conserved dissipative dynamics, i.e.,

$$\frac{\partial \eta_j}{\partial t} = -M_{\eta_j} \frac{\delta \mathcal{F}^{cg}}{\delta \eta_j^*} + \varsigma_{\eta_j}, \quad (j = 1,2,3) \tag{8}$$

$$\frac{\partial \bar{\psi}}{\partial t} = \nabla \cdot \nabla \left(M_{\bar{\psi}} \frac{\delta \mathcal{F}^{cg}}{\delta \bar{\psi}}\right) + \nabla \cdot \varsigma_{\bar{\psi}}, \tag{9}$$

where $M_{\eta_j}$ and $M_{\bar{\psi}}$ represent mobility parameters of the amplitude $\eta_j$ and $\bar{\psi}$, respectively. Besides, coarse grained thermal fluctuations for $\eta_j$ and $\bar{\psi}$ could be considered by appending the stochastic variables $\varsigma_{\eta_j}$ or $\varsigma_{\bar{\psi}}$ to the right hand of Eq. (8) or (9), which are not concerned in this work.

In terms of the transport properties arising from the spatially heterogenous dynamics, Eq. (8) and (9) could well describe the translational motions of the crystal ordering. Only evolutions of the complex amplitudes can describe the rotational motions of the ordering, but Eq. (8) merely grasp the ones resulting from the slow diffusions of $\bar{\psi}$ and $\eta_j$. Namely, the rotational motions are only partly grasped. To seek a complete description of the spatially heterogenous dynamics, fast rotational motions resulting from the elastic relaxations should be properly formulated.

In the classical density functional theory, description of the elastic phonon relaxations is missing. It is found recently by us that the effects of elastic relaxations could be well described by allowing for kinetic motions of the potential lattice. Motions of the reciprocal lattice vectors ($\mathbf{K}_j$) satisfy $\partial \mathbf{K}_j / \partial t = M_{\mathbf{K}_j} \widetilde{\mathbf{p}}_j$, where $\widetilde{\mathbf{p}}_j$ is the partial force along $\mathbf{K}_j$, and $j = 1, 2, ..., N$. We further demonstrate that such equation could be well approximated by a surprisingly simple motion equation obeyed by the phase of the kinetic factor (i.e., $\varphi_j = \mathbf{K}_j \cdot \mathbf{a}_j$, $\mathbf{a}_j$ is the *j*-th primative lattice vector), that is

$$d\varphi_j/dt = \mathbf{a}_j \cdot \partial \mathbf{K}_j/\partial t = |\eta_j|^2 \left(1 + c_h |\eta_j|^2\right) \varsigma, \tag{10}$$

where $c_h = 4\pi$, $\varsigma = \pi\zeta$ and $\zeta$ is a stochastic variable within [-$\zeta_{max}$, $\zeta_{max}$]. Despite of some influences on numerical stabilities, various $\zeta_{max}$ within (0,0.01] does almost not affect the simulation results.

If the spatial variations of the RLVs are permitted, lattice translation invariance should be considered when reconstructing $\psi$ (See Eq. 6). Supposing that $\mathbf{a}_j$ is the *j*-th primitive lattice vector relating to the *j*-th initial RLV ($\mathbf{K}_j^0$) by $\mathbf{a}_j \cdot \mathbf{K}_j^0 = 2\pi$. When RLV ($\mathbf{K}_j$) deviates from its initial value, the product between $\mathbf{a}_j$ and $\mathbf{K}_j$ will departure from $2\pi$. Alternatively, we could fix the RLV at its initial value while change $\mathbf{a}_j$. Denoting the change of $\mathbf{a}_j$ by $\tilde{\mathbf{u}}_j$, the following products should be equal:

$$\mathbf{K}_j \cdot \mathbf{a}_j = \mathbf{K}_j^0 \cdot (\mathbf{a}_j + \tilde{\mathbf{u}}_j) \text{ or } \mathbf{K}_j^0 \cdot \tilde{\mathbf{u}}_j = \mathbf{K}_j \cdot \mathbf{a}_j - 2\pi. \tag{11}$$

Note that $\tilde{\mathbf{u}}_j$ represents extra displacement due to nonequilibrium mechanical relaxations and should be considered during reconstructions of $\psi$. Thereby, Eq. (6) becomes

$$\psi(\mathbf{x}) = \sum_{j=1}^{N} \eta_j \exp[i\mathbf{K}_j^0 \cdot (\mathbf{x} - \tilde{\mathbf{u}}_j)] + c.c.$$

$$= \sum_{j=1}^{N} \eta_j' \exp(i\mathbf{K}_j^0 \cdot \mathbf{x}) + c.c., \tag{12}$$

with the aids of Eq. (11), where $\eta_j'$ is defined by

$$\eta_j' \equiv \eta_j \exp(-i\mathbf{K}_j \cdot \mathbf{a}_j) = \eta_j \exp(-i\varphi_j). \tag{13}$$

Because the original PFC model does not allow for variations of RLVs, we have replaced the RLVs in Eq. (6) with its initial value during above derivations. Eq. (12) indicates that evolutions of the reduced density will incorporate contributions from the nonequilibrium relaxations if multiplying $\eta_j$ by $\exp(-i\varphi_j)$ every time after $\eta_j$ and $\bar{\psi}$ are solved by Eq. (8) and (9), respectively. Hereafter, we refer to such model as kinetic amplitude-expanded PFC (KAPFC) model for brevity. In comparison with the original PFC model, the key ingredient of the KAPFC model is that it allows for additional rotational motions of the crystal ordering arising from the elastic relaxations. The additional rotational motions only rely on the microstructures of the current configuration evolving according to Eq. (8) and (9), while the translational and rotational motions are controlled by $M_{\bar{\psi}}$ and $M_{\eta_j}$, respectively. If not specified, $M_{\bar{\psi}} = M_{\eta_j} = 1$ will be adopted. With the complete descriptions of the two basic motions, plausible results for the nonequilibrium crystallizations of supercooled liquids are obtained through the KAPFC simulations and the details are shown in the Part III and Part IV.

A finite element code with adaptive mesh techniques is implemented for the KAPFC model. Benefiting from the "fast" dynamics as well as the high numerical tolerances of in minimal grid size and maximum timestep, the KAPFC model allows us to approach microstructure evolutions in very large crystals (up to micro-scale) with key atom details. Moreover, interactions among crystallites during nonequilibrium crystallizations are incorporated implicitly in this model.

### III. NONEQUILIBRIUM SOLIFICATIOINS OF HEXAGONAL CRYSTALS

Considering the thermodynamic equilibrium condition of $\epsilon = 37\bar{\psi}^2/15$, we fix initial $\bar{\psi}$ at -0.2

and explore several $\epsilon$ ranging from 0.1 to 0.2 to check the roles of undercooling played on the dendrite growth of hexagonal phases. Four hexagonal samples with periodic conditions applied along X and Y directions are employed to investigate the effects of crystal seed distribution and its initial rotation angle ($\theta$). To this end, nine seeds with different rotation angles uniformly distribute over the first sample (I), one seed with $\theta = 0°$ is placed on the center of the second sample (II), two seeds with $\theta = 0°$ are placed on the center and the conners of the third sample (III) and two seeds with $\theta = 0°$ and $\theta = 15°$ are uniformly placed along the X axis of the fourth sample (IV). The influences of the seed spacing are examined by adjusting the dimension ($L_x \times L_y$) of the samples. For the first three samples, the explored dimensions ranges from $1024a_{hex} \times 512\sqrt{3}a_{hex}$ to $4096a_{hex} \times 2048\sqrt{3}a_{hex}$. The dimension of sample IV is $4096a_{hex} \times 1024\sqrt{3}a_{hex}$ where the distribution of seeds is comparable to the one of sample II (See Fig. 1). The lattice parameter $a_{hex}$ is $4\pi/\sqrt{3}$ in dimensionless unit. Without losing generality, initial unrotated amplitudes are assumed to be real and equal in magnitude whose value is obtained to be $\eta_0 = \left(-3\bar{\psi} \pm \sqrt{15\epsilon - 36\bar{\psi}^2}\right)/15$ through minimizing the free energy. The minimal grid size is $a_{hex}$ and timestep ranges from 0.1 to 0.25 in dimensionless unit depending on the value of $\epsilon$.

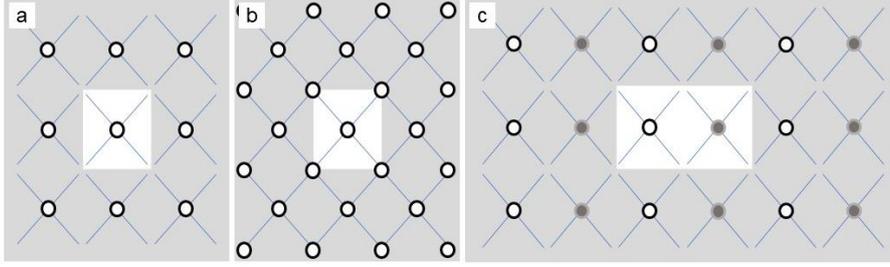

Fig. 1. Initial sites of crystallites in (a) sample II, (b) III and (c) IV, where the white region is the actual domain simulated because of the periodic boundary conditions. The white (gray) circle denotes seed with $\theta = 0°$ (15°) and the blue grid shows the background lattice symmetry, i.e., hexagon (Only two edges are shown for concision). Note that the distribution of the sample I and IV are the same except for the seed types. Overall, the seeds in sample II and IV are nearly square-distributed (Because of the constraints of the lattice symmetry and the periodic boundary conditions, strict square distribution is impossible.), while the seeds in the sample III are hexagon-distributed.

### A. Nonequilibrium patterns during crystallizations of supercooled liquids

Fig. 2b shows the polycrystal growth morphonology in the sample I at moderate undercooling where the spherulites emerging in our simulations extremely resemble with the experimental ones (Fig. 2a). Different colors in Fig. 2b represent the distortion degree of the local lattice in the growing crystals. Alternatively, the local lattice distortions could be observed more directly from Fig. 2c where curved stripes in the growing crystal grains signalize the distortions. The distortions arise from the stress field stimulated by dislocations as well as the boundaries, including the grain boundaries and liquid-solid interface, in the grains. With the growing undercooling, the patterns of the growing grains turn out to be complex morphonology before forming polycrystals (See *Supplementary Materials*).

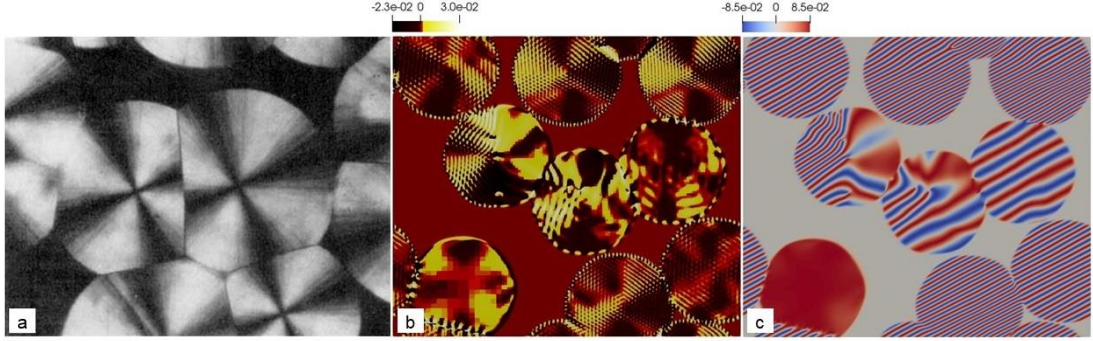

Fig. 2. (a) Melt crystallized spherulites in random block copolymer TMPS/DMS (50/50 wt%) at 30 °C (micrographs reviewed by Magill[3]). (b) and (d) are spherulites crystallized at t = 1800 in the sample I with $\epsilon = 0.125$. The field plotted in the figure b are shear stress, defined by $(\sigma_{11} - \sigma_{22})/2$ where $\sigma_{ij}$ ($i,j$ = 1, 2) is the stress component. While the field plotted in the figure c are $\text{Re}(\eta_1)$. Other simulation parameters are: $\bar{\psi} = -0.2$ and $\zeta_{max} = 5\times10^{-7}$.

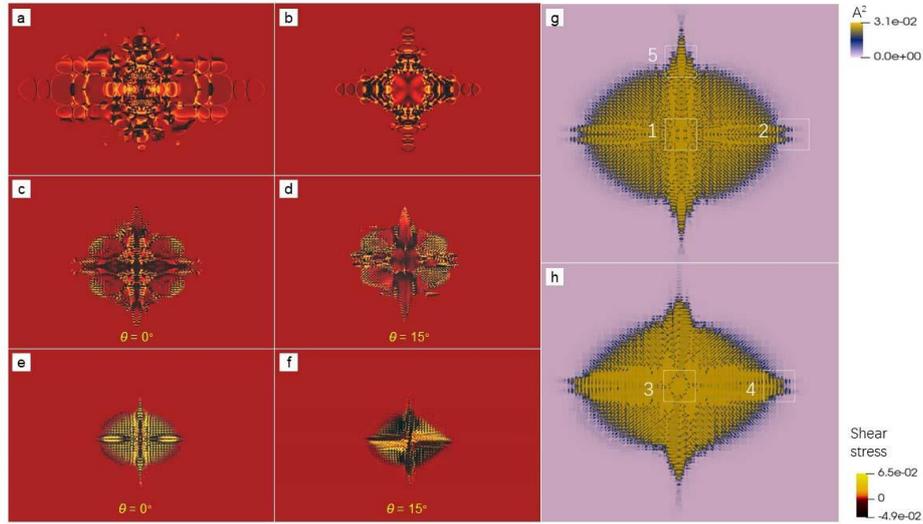

Fig. 3. Morphonology of crystal growth in the different samples with various undercoolings. (a) is columnar dendrite in the sample II with $\epsilon = 0.125$. (b) is equiaxial dendrite in the sample III with $\epsilon = 0.125$. (c-d) are equiaxial dendrites in the sample IV with $\epsilon = 0.125$ and (e-f) are the same but with $\epsilon = 0.15$. Coloring scheme for (a-f) are the same as that in Fig. 1c-d. Lattice defects (mainly dislocations) in (e) and (f) can be observed more clearly from the $A^2$ field plots, i.e., (g) and (h), where five regions, marked by the white squares, size of 500×500 for each. From the figures, dislocations are shown by the blue color.

To make clear of the pattern formation mechanism, we resort to the growth kinetics of "single" seed. Actually, the seed is still in a polycrystal but with much larger seed spacing because of the periodic conditions. The results are shown in the Fig. 3. Surprisingly, at moderate undercooling ($\epsilon = 0.125$), the CET takes place when the seed distribution changes from square symmetry (Fig. 3a) to hexagonal symmetry (Fig. 3b), which is not predicted by traditional models. Such CET arises from the combined effects of the undercooling and lattice symmetry. The columnar dendrite is prone to growing along X axis of the sample I, while the equiaxial dendrite grows simultaneously along X and Y axis of the sample

II. Keeping the square symmetry but rotating the right (as well as the left) seed by 15°, the columnar dendrite becomes more complex equiaxial dendrite. Except for the four primary arms growing along X and Y directions (See Fig. 2c-d), multiple side arms are observed in the equiaxial dendrite. Further increasing the undercooling, the equiaxial dendrite turn out to be even slenderer (See Fig. 2e-f). Such phenomenon results from the large disturbances around the solid-liquid interface caused by the elastic interactions. Interestingly, the growth tip of the left (or right) primary arm is split (See Fig. 3c, d, g, h), resembling with the tip-splitting of dendrites arising from heterogeneous interface energy [28, 29]. This result indicates that interactions among seeds with different types could affect the interface energy. Overall, the envelope of the equiaxial dendrite is rectangular (or ellipsoid) rather than square (or circle) during the growth. This is because the seeds are not strictly square-distributed (See Fig. 1). Dislocation density is much larger in regions between the primary arms (See Fig. 2g-h) due to the impinging of small grains formed by the frequent occurrence of the GFN.

Typical local microstructures of the equiaxial dendrites at atom scale are reconstructed by applying Eq. (2) and the results are shown in Fig. 4. Dislocation configurations at atom level could be clearly identified from the results (Fig. 4a and c). Interestingly, dislocations mainly emerge outside a circular region centered at the seed (such as the region "1" and "3" in Fig .2g-h). After sufficient relaxation time, these dislocations would partly annihilate accompanied by rotation and shrinking of the small grains. Besides, the GFN at the growth tip and the dendric growth with side arms are observed (See Fig. 4b, d-e). According above discussions, we have reproduced many typical nonequilibrium patterns, including spherulites, columnar dendrite and equiaxial dendrite, during crystal growth by the KAPFC model. It is found that, except for the undercooling, the CET also depends on the seed distributions and lattice symmetry, which is not aware of previously. Dislocation density notably increases only after a certain period of the crystal growth and would become larger with the promoting undercooling.

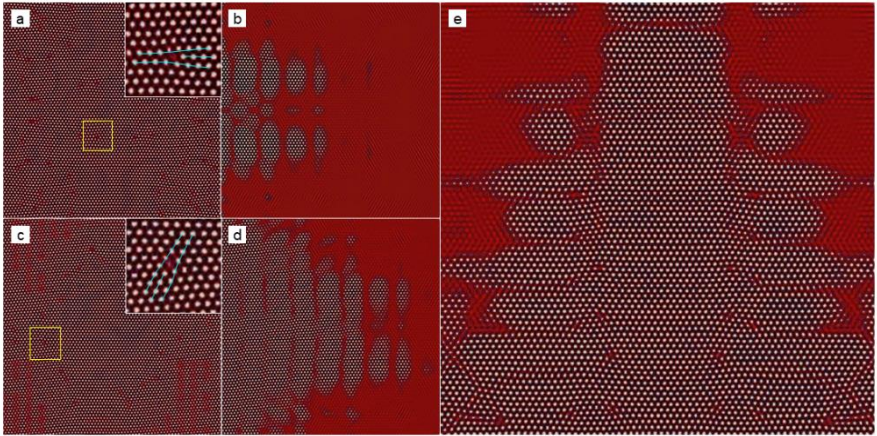

Fig. 4. Reconstructed atom density field of the five regions in Fig. 2g and 2h. (a-e) correspond to the region numbered from "1" to "5", respectively. The inset in the figure a (or c) is the magnified view of the corresponding region marked by the yellow square, where the core structures of the dislocations are clear seen.

**B. Kinetics of the growth fronts**

The kinetics of the right (upper) growth front for the sample II are shown in Fig. 5a (b). Two growth stages could be identified. In the first growth stage, corresponding to the growth stage I in Fig. 5a, crystal grows mainly through diffusive transport of mass, i.e., slow mode, whose interface velocity ($v$) vs growth time ($\tau = t\text{-}t_0$, $t_0$ is the nucleation time) characterized by the relation of $v \propto \tau^{-1/2}$. This result agrees with the one [30] predicted by the diffusive PFC model except for larger velocity coefficient because of the higher undercooling. Density of lattice defects characterized mainly by dislocations is small in this stage. As results, a relatively "clean" region centered at the seed is observed in the growing crystal (See Fig. 3). With the growth of the clean region, local undercooling at growth front increases because of the low density caused by the increasing depletion layer. The emergence of the low-density layer arises from the increment of the average density in the solid. As the solid grows, the low-density layer at the growth front extends more and more deep into the liquid. Its extending speed is proportional to the bulk wave speed of the liquid that are much fast than the diffusive transport processes. If the low-density perturbations are sufficiently large, new crystallites would nucleate and growth, which in turn generate new depletion layer perturbations superpositioned on the old one. The superposition effects, on the one hand, accelerate the merging of existing crystallites and, on the other hands, promote the GFN and then repeat such processes again. It is such reason that makes the GFN-dominated growth processes, i.e., the growth stage II in the Fig. 5a, begin. A notable feature in this stage is that the growth front moves forward through a combined mechanism of the diffusion-controlled anisotropic growth and the GFN-controlled growth, which leads to the step-growth style of the front (See Fig. 5a). Within each "step", crystal grows still through the diffusion-controlled mechanism. In contrast to the steady growth speed observed in solid-liquid coexistence regime from both PFC simulations[12] and experiments on colloidal hard sphere crystallization[31], the interface velocity vs growth time asymptotically approaches to a relation of $v \propto \tau^{n-1}$ at sufficient growth time, where $n \sim 2.0$ in present work (See Fig. 5). We further investigate the crystal growth in the sample II under different undercooling. The two stages are also observed at lower or higher undercoolings (See *Supplementary Materials*).

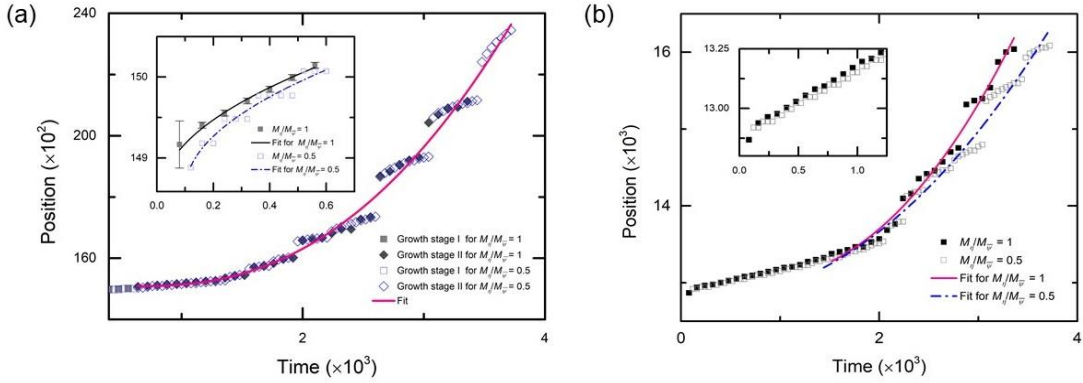

Fig. 5. (a) Right-growth-front position versus time for the sample II with $\epsilon = 0.125$, where the gray solid line (or blue dot-dash line) for growth stage I in the inset is a fit to relation[30]: $Z = Z_0 + C\sqrt{t - t_0}$, $Z_0 = 14858.64$, $t_0 = 20.00$ and $C = 6.59$ (or $Z_0 = 14861.66$, $t_0 = 102.06$ and $C = 6.60$), and the pink solid line for the growth stage II ($M_\eta/M_{\bar\psi} = 0.5$) is a fit to $Z = Z_0 + C(t - t_0)^n + Ae^{-(t-t_0)/\tau_0}$, $Z_0 = 14262.90$, $C = 5.15 \times 10^{-5}$, $n = 2.36$, $t_0 = 638.72$, $A = 828.90$ and $\tau_0 = 2.92 \times 10^3$. (b) Upper-growth-front position versus time for the sample II with $\epsilon = 0.125$, where the pink solid line (or blue dot-dash line) is a fit to the same relation satisfied by the stage II for the case of $M_\eta/M_{\bar\psi} = 1$ (or $M_\eta/M_{\bar\psi} = 0.5$) and the fitting parameters are $Z_0 = 12997.78$, $C = 3.78 \times 10^{-5}$, $n = 2.30$, $t_0 = 539.89$, $A = 0.16$ and $\tau_0 = 1.70 \times 10^2$ (or $Z_0 = 12747.83$, $C = 6.58 \times 10^{-5}$, $n = 2.17$, $t_0 = 11.37$, $A = 4.88$ and $\tau_0 = 1.41 \times 10^2$). The inset is the magnified view of the first growth stage.

Specially, to examine the role of the ratio of $M_\eta/M_{\bar{\psi}}$ on the crystal growth, we modify the ratio of $M_\eta/M_{\bar{\psi}}$ by setting $M_{\bar{\psi}} = 2.0$. Results for the kinetics of the right (upper) growth front are shown in the Fig. 5. The two growth stages also emerge, but are slightly differed in the quantized manner. Larger diffusion mobility of the average density slightly promotes the growth speed along X direction at the first stage while hinders the growth along Y direction at the second stage. This result implies that the larger diffusion mobility of the average density benefits the formation of columnar dendrites. The transition time between the two growth stages relates to the undercooling by a power law (See Fig. 6). Such result suggests that the second growth stage would emerge in the crystallization system with either large seed spacing and low undercooling or smaller seed spacing and high undercooling. By controlling the transition time through the undercooling and the seed spacing (related to the system size, impurity content among others), polycrystals with different grain size, grain shape and dislocation density are ready to be acquired.

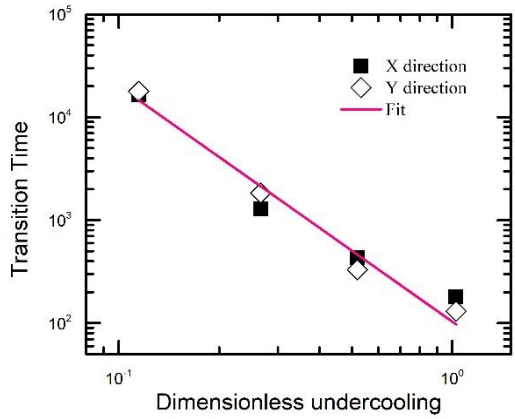

Fig. 6. Transition time between the growth stage I and II versus undercooling reduced by the equilibrium $\epsilon$. The pink solid line is a fit to a power law, satisfying $t_{I\leftrightarrow II} = 103.58 * (\epsilon/\epsilon_0)^{-2.29}$.

### C. Microscopic mechanism of the crystal growth

Typical scenarios of the columnar dendrite growth at atom scale are shown in the figure Fig. 7. Fig. 7a-b illustrates the diffusion-controlled stage, which is featured by the faceting growth. In this stage, the interface moves forwards layer by layer along the direction perpendicular to the closely packed direction of the hexagonal phase. Due to the large undercooling at the growth front, the depletion layer extends deeply into liquid and cause nucleation of new crystallites in the front of the interface (See Fig. 7c), which is the notable feature of the second growth stage. New crystallites would quick merge with the parent one and leave grain boundaries (or dislocations) between them because of their different orientations (See Fig. 7d-e, j-k). Then the depletion layer perturbations of both old and new crystallites would trigger new nucleation in an extremely complex form (See Fig. 7f-i). When crystallite number are more one, the growths of these crystallites at this stage are inevitably coupled. As results, various morphonology of crystal growth would form. Similar growth processes are also observed in the other samples, such as sample II (See *Supplementary Materials*). The growth morphology mainly depends on the undercooling and the seed distribution. Overall, the two growth stages generally exist during the nonequilibrium crystallization and the complex growth morphonology mainly formed through the GFN mechanism in the latter stage. The transition time between the two stages decreases rapidly with the growing undercooling. Thus, the second growth stage could emerge in small crystals (with small seed spacing) when the undercooling is sufficiently high. Because the high dislocation density generating in the second growth stage also increases with the growing undercooling, especially at the early time of the growth stage, amorphous nucleation may happen.

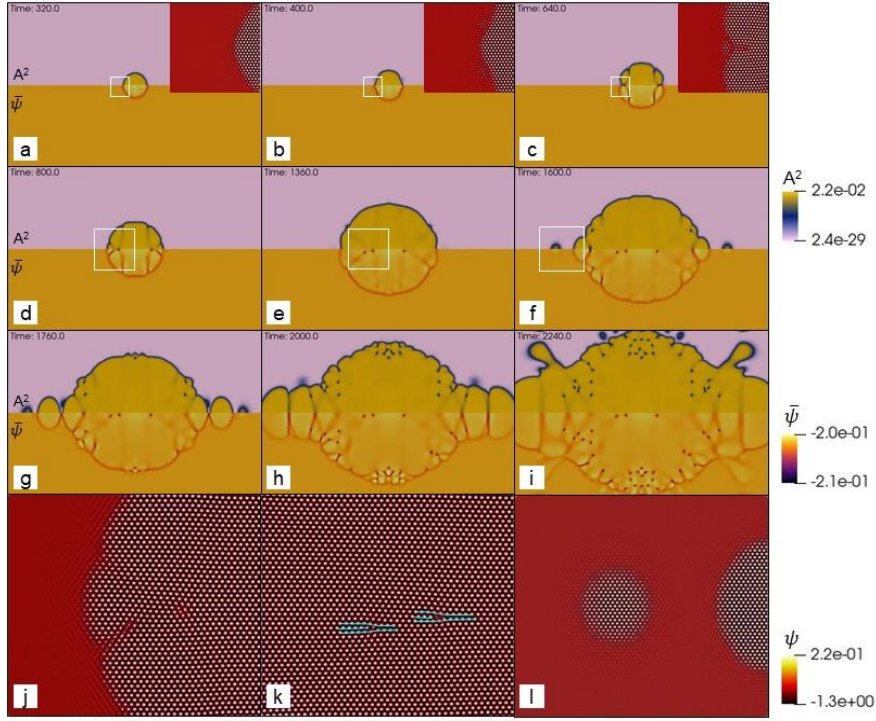

Fig. 7. Enlarged contour plots at different moments showing the columnar dendrite growth processes of the hexagonal phase at $\epsilon = 0.125$. In figure (a-i), the upper panel plotted is field $A^2$ while the lower is $\bar{\psi}$. Reconstructed atom-number-density fields of the region marked by the blue rectangles in figure (a-f) are given in the inset of (a-c) and in the figure (j-l) for (d-f), respectively. Although initial simulated system is centrosymmetric, dislocations generated during the crystal growth may slightly deviate from the symmetric distribution because of the random driven force for the RLV evolutions.

## IV. FACETING AND DENDRITIC GROWTH OF BCC CRYSTALS

To further confirm our results obtained from the 2D lattice, large scale KAPFC simulations are also conducted to investigate the crystal growth in BCC lattice. A small BCC seed with [100], [010] and [111] aligning along X, Y and Z axis, respectively, are placed at the center (i.e., the origin point) of a simulation box initially filled with liquid phase. The size of the simulation system is $80a_{bcc} \times 80a_{bcc} \times 80a_{bcc}$, where dimensionless lattice parameter $a_{bcc}$ is $2\sqrt{2}\pi$. Periodic boundary conditions are applied along X and Y directions. Model parameters $(\epsilon, \bar{\psi})$ are selected to be (0.35, -0.35) and (0.40, -0.35), corresponding to low and high undercooling in the phase diagram[32]. Assuming that the initial amplitudes are real and equal in magnitudes (represented by $\eta_0$), we get $\eta_0 = \left(-2\bar{\psi} \pm \sqrt{5\epsilon - \bar{\psi}^2}\right)/15$ after minimizing the free energy functional. The minimal grid size is $1.25a_{bcc}$. Timestep is 1.0 in dimensionless unit.

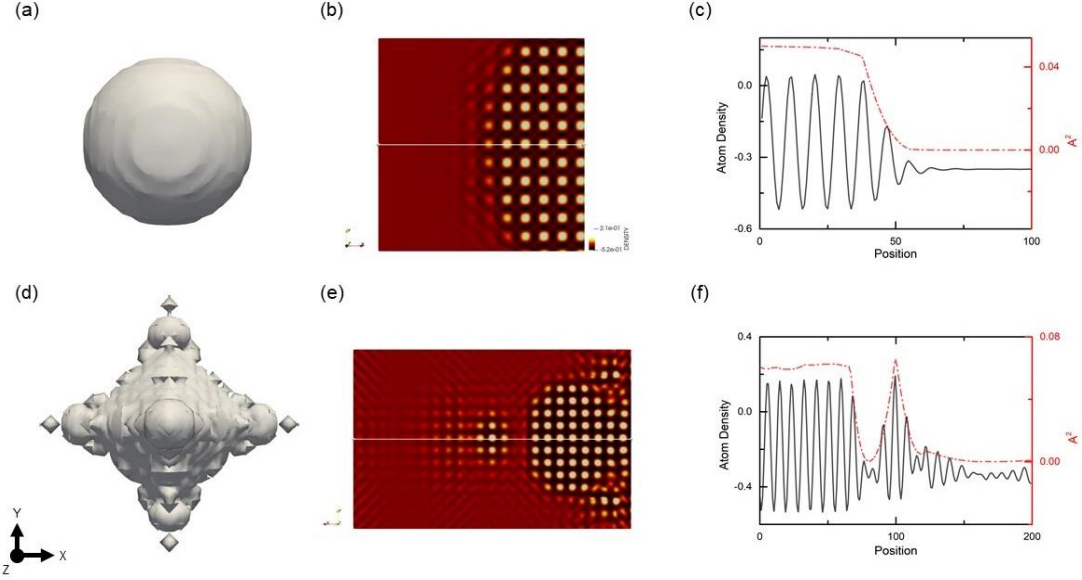

Fig. 8. Faceting morphonology and equiaxial dendrite in the BCC sample (a-c) at t = 260 with $\epsilon = 0.35$ and (d-f) at t = 180 with $\epsilon = 0.4$, respectively. The contour surfaces in the figure (a) and (d) correspond to $A^2 = 0.01$. In the figure (b) (or e) are the enlarged reconstructed-density view of the plane Z = 0 (i.e., (001) plane) at the left (right) growing tip. Figure (c) (or f) are the profiles of atom density and $A^2$ along the white line drawn in the figure (b) (or e). For sakes of comparison, the density profile is drawn along the line from the solid to the liquid.

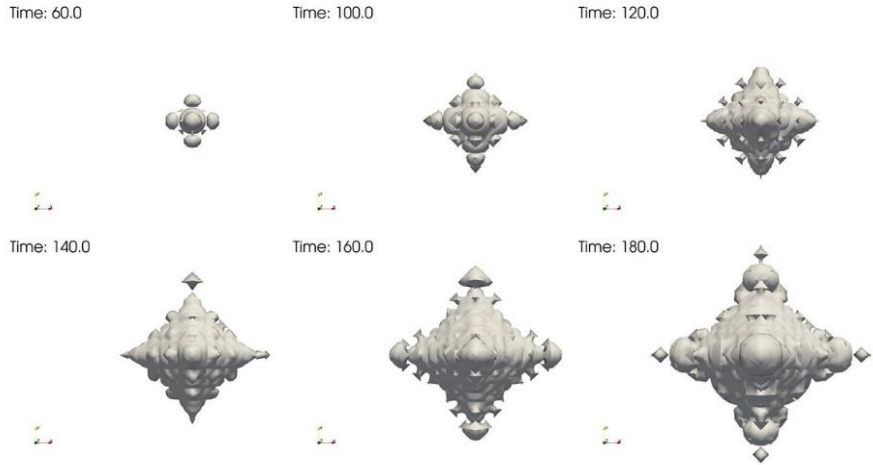

Fig. 9. Dendritic growth of the BCC crystal through the GFN mechanism. The contour surfaces are the same as the ones in the figure Fig. 8.

The results are shown in Fig. 8, where faceting morphonology and equiaxial dendrite are observed. The growth fronts for the two cases (corresponding to the two growth stages) are featured by clean and smearing solid-liquid interfaces, respectively (See Fig. 8b, c, e, f). At the low undercooling, the diffusion-controlled growth stage covers the whole simulated crystallization process. When the simulation time (as well as the seed spacing) is sufficiently long (large), the second growth stage would emerge according to the results in Part IIIc. Alternative, we could observe such growth stage through elevating the undercooling with less computational efforts. The results at the high undercooling are shown in Fig. 8d-f and Fig. 9. It is found that the GFN-dominated growth quick takes over the crystallization controlled by the mass diffusion and results in formations of the complex nonequilibrium pattern, i.e., the equiaxial dendrite, which confirms our assertions above. In the real crystallization system, the seed spacing is usually much larger than the one explored in present work. This will lower the undercooling condition required to observe the second growth stage.

## V. SUMMERING AND CONCLUSIONS

In summary, the nonequilibrium crystallizations are investigated using a minimal extended PFC model, i.e., the so-called KAPFC model. In contrast to other variants of the PFC model, the KAPFC model proposed in present work could grasp the elastic phonon relaxations missing in the original PFC model with least extension where all conceptions employed are from the original model and thus no reinterpretations among conceptions at different levels are needed. Another remarkable advantage of the model is the high computational efficiency due to the "fast" dynamics as well as the high numerical tolerances of in minimal grid size and maximum timestep. We employ such model to investigate the formations of the nonequilibrium patterns in the hexagonal lattice, which typically requires large simulation systems and long simulation time hardly accessed by other method operated at atom level. Nonequilibrium patterns, including the faceting growth, spherulite, dendrite and the CET among others at atom scale, are revealed unprecedently. It is found that the formations of the nonequilibrium patterns depend on the lattice symmetry, undercooling and interactions among the growing crystallites. Particularly, roles of the interactions arising from seed distributions as well as seed types (characterized by initial rotation angular) are investigated. The result indicates that the interactions could result in the CET as well as tip-splitting of the dendrites, which are not predicted before. Through examining the kinetics of growth front under various undercooling, two growth stages, i.e., the diffusion-controlled anisotropic growth and the GFN-dominated growth, are identified. At large undercooling, the diffusion-controlled growth stage become so short that it is difficult to be aware of its existence compared with the second stage. Dislocations are dramatically generated in the second stage mainly through impinging of small crystallites formed via the GFN mechanism. This may be the source for the precursor of amorphous nucleation[14]. Nonequilibrium solidifications of BCC crystals are also investigated, which confirms above conclusions. The result of present work provides clues for designing the polycrystals containing gains with various shapes and sizes, as well as different initial dislocation densities, through controlling the undercooling and the seed distribution, to meet the various requirements of application realms.

## ACKNOWLEDGMENTS

This work is financially supported by the National Key Research and Development Program of China (No. 2018YFB0704000), National Natural Science Foundation of China (NSFC-NSAF U1830138, 12072044), the Fundamental Research Funds for the Central Universities and National Defense Science and Technology Key Laboratory Foundation (No. 6142A05200102).

**APPENDIX: DETAILED MOTION EQUATIONS FOR HEXAGONAL AND BCC LATTICES**

The KAPFC model employed in present work is an extension of the amplitude-expanded phase field crystal model[24, 27]. For hexagonal lattice, the final expression for Eqs. (8-9) in the main text are

$$\frac{\partial \eta_j}{\partial t} = -M_{\eta_j}\left\{\mathcal{G}_j^{\,2}\eta_j - (\epsilon - 3\bar{\psi}^2)\eta_j + 3\left(2A^2 - |\eta_j|^2\right)\eta_j + 6\bar{\psi}\prod_{\substack{i=1\\i\neq j}}^{N}\eta_i^*\right\} + \varsigma_{\eta_j}, \tag{A1}$$

$$\frac{\partial \bar{\psi}}{\partial t} = M_{\bar{\psi}}\nabla^2\left[(-\epsilon + 1)\bar{\psi} + \bar{\psi}^3 + 6\bar{\psi}A^2 + 6\left(\prod_{j=1}^{N}\eta_j + \prod_{j=1}^{N}\eta_j^*\right)\right] + \nabla \cdot \varsigma_{\bar{\psi}}. \tag{A2}$$

The KAPFC model specially considered the effects of local stress relaxations in terms of RLV evolutions which turns out to a simple equation, i.e., Eq. (10) in the main text. Rigors derivations as well as numerical verifications are given in our another work (in preparation). This model could not only reproduce various well-known nonequilibrium patterns during the crystal growth, but also correctly describe the classical phenomenon —— the rotation and shrinking of circular grains embedded in the hexagonal lattice[17, 33].

For BCC lattice, the RLVs are chosen to be

$$\mathbf{K}_1 = k_0(1,1,0), \quad \mathbf{K}_2 = k_0(1,0,1), \quad \mathbf{K}_3 = k_0(0,1,1),$$
$$\mathbf{K}_4 = k_0(0,1,-1), \quad \mathbf{K}_5 = k_0(1,-1,0), \quad \mathbf{K}_6 = k_0(-1,0,1). \tag{A3}$$

where $k_0 = \sqrt{2}/2$. Note that triadic and quartic resonances are satisfied, i.e.,

$$\mathbf{K}_1 - \mathbf{K}_4 - \mathbf{K}_2 = 0, \quad \mathbf{K}_1 + \mathbf{K}_6 - \mathbf{K}_3 = 0, \quad \mathbf{K}_4 + \mathbf{K}_5 + \mathbf{K}_6 = 0, \quad \mathbf{K}_2 - \mathbf{K}_5 - \mathbf{K}_3 = 0, \tag{A4}$$

and

$$\mathbf{K}_1 - \mathbf{K}_3 - \mathbf{K}_4 - \mathbf{K}_5 = 0, \quad \mathbf{K}_1 - \mathbf{K}_2 + \mathbf{K}_5 + \mathbf{K}_6 = 0, \quad \mathbf{K}_2 - \mathbf{K}_3 + \mathbf{K}_4 + \mathbf{K}_6 = 0. \tag{A5}$$

Instead of applying Eqs. (A1-2), the motion equations of the amplitudes and the average density for the BCC lattice are

$$\frac{\partial \eta_j}{\partial t} = -M_{\eta_j}\left\{\mathcal{G}_j^{\,2}\eta_j - (\epsilon - 3\bar{\psi}^2)\eta_j + 3\left(2A^2 - |\eta_j|^2\right)\eta_j + \mathcal{Q}_j\right\} + \varsigma_{\eta_j}, \tag{A6}$$

$$\frac{\partial \bar{\psi}}{\partial t} = M_{\bar{\psi}}\nabla^2[(-\epsilon + 1)\bar{\psi} + \bar{\psi}^3 + 6\bar{\psi}A^2 + 6(\eta_1^*\eta_2\eta_4 + \eta_3^*\eta_1\eta_6 + \eta_4\eta_5\eta_6 + \eta_2^*\eta_3\eta_5 + c.c.)] + \nabla \cdot \varsigma_{\bar{\psi}} \tag{A7}$$

where

$$\mathcal{Q}_j = -(g - 2\lambda\bar{\psi})(\eta_k\eta_n^* + \eta_j\eta_l) + 2\lambda(\eta_k\eta_l\eta_m + \eta_j\eta_m^*\eta_n^*), \quad (j = 1,2,3) \tag{A8}$$

$$\mathcal{Q}_l = -(g - 2\lambda\bar{\psi})(\eta_m^*\eta_n^* + \eta_i\eta_j^*) + 2\lambda(\eta_i\eta_k^*\eta_m^* + \eta_k\eta_j^*\eta_n^*), \quad (l = 4,5,6) \tag{A9}$$

where $(i,j,k)$ or $(l,m,n)$ is a permutation of $(1,2,3)$ or $(4,5,6)$. Combined Eqs. (A1-2) or Eqs. (A6-7) with Eqs. (10, 13), the amplitudes as well as the average density at each moment could be obtained.